\documentclass[12pt,a4,notitlepage]{article}

\usepackage{amssymb}
\usepackage[mathscr]{eucal}
\usepackage[intlimits]{amsmath}
\usepackage{theorem}

\newcommand{\fie}{\varphi}
\newcommand{\eps}{\varepsilon}
\newcommand{\heps}{\hat\epsilon}

\newcommand{\Sop}{{\cal S}}

\font\cmbms=eufm10 at 12pt
\newcommand{\specz}{\text{\cmbms{z}}}
\newcommand{\dy}{\frac{\text{d}y}{\text{d}z}}
\newcommand{\ddy}{\frac{\text{d}^2y}{\text{d}z^2}}

\newcommand{\wpsi}{\widehat\psi}

\title{Painlev\'e equations from Darboux chains\\
Part 1: $P_{III}-P_V$} 

\author{Ralph Willox${}^{1,3}$  and Jarmo Hietarinta${}^{2,1}$\\ \\
 ${}^1$Graduate School of Mathematical Sciences, University of Tokyo,\\
  3-8-1 Komaba, Meguro-ku, 153-8914 Tokyo, Japan.\\
${}^2$Department of Physics,
University of Turku\\ FIN-20014 Turku, Finland.\\
${}^3$Theoretical Physics, Free University of Brussels (VUB),\\
Pleinlaan 2, 1050 Brussels, Belgium.}
\begin{document}

\maketitle

{\abstract We show that the Painlev\'e equations $P_{III}-P_{VI}$ can
  be derived (in a unified way) from a periodic sequence of Darboux
  transformations for a Schr\"odinger problem with quadratic
  eigenvalue dependency. The general problem naturally divides into
  three different branches, each described by an infinite chain of
  equations. The Painlev\'e equations are obtained by closing the
  chain periodically at the lowest nontrivial level(s). The chains
  provide ``symmetric forms'' for the Painlev\'e equations, from which
  Hirota bilinear forms and Lax pairs are derived. In this paper
  (Part 1) we analyze in detail the cases $P_{III}-P_{V}$, while
  $P_{VI}$ will be studied in Part 2.}

\section{Introduction}
The aim of this paper is to provide a bridge between two extremely
fruitful descriptions of the Painlev\'e equations in the context of
integrable systems: the (Darboux) dressing chain approach of the
Russian school \cite{S92,VS93,A94,S99,FSV95,AS00} and the tau-function
based approach of the Japanese school
\cite{O81,JM81,O86a,O86b,O87a,O87b,O87c,NY98a,NY99,N00}.

In the seminal paper \cite{A94} Adler proposed a complete description
of the Painlev\'e equations $P_{II-VI}$ by means of so-called Darboux
chains. These generalize the usual dressing chain \cite{S92} which, in
its periodic case, was already known to provide a description of the
$P_{IV}$ and $P_V$ equations \cite{VS93}.  Adler succeeded in
describing the Schlesinger-B\"acklund transformations and notably the
Weyl-group action of these transformations for the Painlev\'e
equations, merely by starting from the Darboux transformations that
underlie their associated (periodic) chain equations. However, its
relationship with Okamoto's theory of the Painlev\'e equations (in
which Weyl-groups play a central role) has still not been elaborated.
In the Okamoto description, the Weyl-group action of the B\"acklund
transformations for the Painlev\'e equations is closely related to the
notion of a {\em tau-function} \cite{O81,O86a,O86b,O87a,O87b,O87c}
(see \cite{O92} for an overview), which in turn is connected to the
existence of isomonodromy problems associated with the Painlev\'e
equations \cite{JM81}. In recent years this (affine) Weyl-group
structure of certain B\"acklund transformations for the Painlev\'e
equations, together with the tau-function description of the
Hamiltonians underlying these equations (see \cite{O99} for a survey
of results), has given rise to a ``symmetry-based'' approach to the
Painlev\'e equations and related systems \cite{NY98a,NY98b}. In
\cite{NY98a} Noumi and Yamada proposed a systematic description of a
class of dynamical systems comprising the $P_{IV}$ and $P_V$
equations, which all possess B\"acklund transformations that make up
an affine Weyl-group and which allow for tau-function descriptions of
their Hamiltonian structures.  So called ``symmetric forms'' of these
dynamical systems play a central role in this approach \cite{N00}.

We will show, using the associated Hirota bilinear representations,
that periodic closings of the standard dressing chain considered in
\cite{VS93,A94,AS00} -- and from which $P_{IV-V}$ are obtained at
periods 3 and 4 respectively -- correspond exactly to the dynamical
systems introduced in \cite{NY98a}.  Our approach is, however, not
restricted to the $P_{IV-V}$ equations or similar cases. We shall also
explicitly derive a symmetric form of the $P_{III}$ equation and of a
whole class of related equations (which first appeared in \cite{S99}),
all of which can be cast into Hirota bilinear form, i.e., their
solutions can always be expressed in terms of tau-functions.

A major part of this paper is devoted to a systematization of the
results obtained by Adler \cite{A94} in order to be able to derive all
relevant Darboux (dressing) chains from a common starting point and to
analyze their properties using the same techniques. This unified
approach then not only pays off when bilinearizing the Darboux chains
we obtain, but also and especially so when deriving Lax
representations of those chains.  Although a lot of attention has been
devoted to the spectral properties of the linear equations whose
Darboux transformations underlie the various (Darboux) dressing chains
associated to the Painlev\'e equations \cite{VS93,DS94,AS00}, no Lax
representations of these chains have been offered in the literature
pertaining to Darboux chains, although the Noumi-Yamada symmetry
approach does yield such representations.  In the following we shall,
systematically, derive Lax pairs for the various Darboux chains
presented throughout this paper. In the $P_{IV-V}$ case these Lax
pairs will turn out to be identical to those obtained in
\cite{N00,T02}, but we believe the result concerning $P_{III}$ is new.

The structure of the paper is as follows: We start by defining the
Darboux transformation for a Schr\"odinger operator with an
energy-dependent potential and derive its associated chain equation.
The analysis naturally divides into three branches covering
$P_{IV}-P_V$, $P_{III}$, and $P_{VI}$, respectively.  In Section 3 we
study the $P_{IV}-P_V$ branch in general and then in Sec.~4 give the
details for $P_{IV}$ and in Sec.~5 for $P_{V}$. Then we study the
$P_{III}$ branch in general and in Sec.~7 give the specific details
for $P_{III}$ itself.  As mentioned above, the results concerning the
$P_{VI}$ case will be discussed in Part 2.

While finishing this paper we came across Ref.~\cite{S99} by Shabat, which
is also concerned with systematizing the Darboux-Painlev\'e
connection. It is therefore not surprising that our analysis runs
largely parallel with that paper and that many formulae are identical
or similar. However, as we wish to focus on the specific connections
to the Painlev\'e equations, continuing with their bilinearization and
Lax pairs, our presentation is both more specific and more detailed.
We believe that in doing so the method and its internal machinery gain
further clarity.

\section{The Darboux transformation}
Roughly speaking, a Darboux transformation is a transformation between
two \hyphenation{li-near}linear (ordinary or partial) differential
equations, of the same overall appearance but for different values of
their coefficients: the solutions of the first equation will be mapped
to the solutions of the second, and the changes in the coefficients of
the equation are also provided by the transformation (see, e.g.,
\cite{MS91} for a classic introduction in the context of integrable
systems).  We are interested in {\em chains} of such transformations
applied to the Schr\"odinger equation.

\subsection{The general form of the chain equations}
Consider the spectral problem associated with the Schr\"odinger
operator \cite{S99}
\begin{equation}\label{E:eig1}
L_j(u,\lambda) \psi_j(\lambda,x) = 0,
\text{ where }\quad L_j(u,\lambda):= \partial_x^2 + u_j(\lambda,x).
\end{equation}
Here $j$ indexes the eigenproblem, and $\lambda$ are the eigenvalues
(assumed non-degenerate).  Note that we make no assumptions about
boundedness or normalizability of the eigenfunctions. We shall refer
to the functions $u_j(\lambda,x)$ as ``generalized potentials'' as
they include dependence on the eigenvalues $\lambda$.
Next define the operator
\begin{equation}\label{E:tra1}
G_j(\lambda,x) :=A_j(\lambda,x)  \left(\partial_x - F_j(\lambda,x)\right),
\end{equation}
and then  using this define new functions
\begin{equation}
\psi_{j+1}(\lambda,x):=G_j(\lambda,x)\psi_j(\lambda,x),\label{E:eigtra}
\end{equation}
for each eigenfunction $\psi_j(\lambda,x)$ of the original problem at
level $j$.  The new functions $\psi_{j+1}$ will then be eigenfunctions
of a new operator $L_{j+1}$ (with the {\em same, generic} eigenvalue
$\lambda$), provided that $L_{j+1}$ satisfies the operator identity
\begin{equation}
L_{j+1}\, G_j(\lambda,x) = \widetilde G_j(\lambda,x) L_j(u,\lambda),
\label{E:comm1}
\end{equation}
for some $\widetilde G_j(\lambda,x) :=A_j(\lambda,x) (\partial_x -
\widetilde F_j(\lambda,x))$. Assuming that $L_{j+1}$ is a differential
operator (i.e., polynomial in $\partial_x$) one finds that
\begin{align}
\widetilde F_j(\lambda,x)&=F_j(\lambda,x)-2
(\log A_j(\lambda,x))',\label{E:Ftil}\\
L_{j+1}&=L_j(u_{j+1},\lambda),\\
u_{j+1}(\lambda,x)&=u_j(\lambda,x)+[2F_j(\lambda,x)A_j(\lambda,x)
-A_j(\lambda,x)']'/A_j(\lambda,x),\label{E:gu}\\
F_j(\lambda,x)'&+F_j(\lambda,x)^2+u_j(\lambda,x)=\mu_j(\lambda)
 A_j(\lambda,x)^{-2},
\label{E:gF}
\end{align}
where $\mu_j$ is an integration constant (the $^\prime$-notation
stands for $\frac{d}{d x}$).

\begin{quote}The transformation \eqref{E:eigtra} from $\psi_j$ to 
  $\psi_{j+1}$ will be called a {\em Darboux transformation} iff at
  each step $j$ the operator $G_j$ is such that {\em it annihilates some
    chosen eigenfunction} $\fie_j$ of \eqref{E:eig1} having eigenvalue
  $\nu_j$.
\end{quote}
This necessarily
implies that $F_j(\nu_j,x) = (\log \fie_j)_x$ and therefore these
$F_j$ also satisfy the equation
\begin{equation}
F_j(\nu_j,x)'+F_j(\nu_j,x)^2+u_j(\nu_j,x)=0.
\label{E:gFd}
\end{equation}
This should be compatible with \eqref{E:gF} and therefore
$\mu_j(\nu_j)=0$.

Now subtracting \eqref{E:gF} for $j$ and for $j+1$ and using
\eqref{E:gu} to eliminate the potentials $u$ we get a chain of
equations
\begin{eqnarray}
F_{j+1}(\lambda,x)'+(F_j(\lambda,x)-(\log A_j(\lambda,x))')'
+\qquad\qquad\qquad&&\nonumber\\
F_{j+1}(\lambda,x)^2-(F_j(\lambda,x)-(\log A_j(\lambda,x))')^2
+\qquad&&\nonumber\\
\mu_j(\lambda)A_j(\lambda,x)^{-2}
-\mu_{j+1}(\lambda)A_{j+1}(\lambda,x)^{-2}&=&0.\label{E:chaing}
\end{eqnarray}
In the following we shall think of this {\em chain equation} as a
generator for $\lambda$-independent chain equations (the {\em dressing
  chains}). Thus the main equations underlying our
\hyphenation{a-na-ly-sis}analysis will be \eqref{E:gu}, which gives
the change in $u$ once $F,\,A$ are given, and \eqref{E:chaing}, the
chain equation.

\subsection{Linear problem for the chain equations}
The commutation relation \eqref{E:comm1} which generates the chain
equation \eqref{E:chaing} can also be rewritten as
\begin{equation}
L_{j+1} G_j - G_j L_j = 2 A_j^\prime L_j\qquad\text{or}
\qquad\quad A_j^2 L_{j+1} G_j = G_j A_j^2 L_j,
\end{equation}
which allows one to interpret the system (\ref{E:eig1},\ref{E:eigtra}) as a
kind of linear problem for the chain equation.  Let us define
\begin{equation}
M_j := \partial_x - A_j^{-1} \Sop - F_j,\qquad\text{where}
\quad \Sop : \psi_j(\lambda,x)\mapsto \psi_{j+1}(\lambda,x),
\label{E:Mop}
\end{equation}
then since 
\begin{equation}
M_j \psi_j(\lambda,x) = 0\label{E:Mlin}
\end{equation}
(i.e., equation \eqref{E:eigtra}) we can eliminate $\partial_x$ from
the eigenvalue problem \eqref{E:eig1} and obtain the following second
order difference equation for the eigenfunctions $\psi_j(\lambda,x)$:
\begin{gather}
L_j^d \psi_j(\lambda,x) = 0,\label{E:Ldlin}\\\intertext{where}
L_j^d := A_j^{-1} \Sop\, A_j^{-1} \Sop + 
A_j^{-1} \Sop\, F_j + F_j\,A_j^{-1}\Sop - (\log A_j)'\, A_j^{-1}\Sop 
+ \mu_j A_j^{-2}\label{E:Ldop}
\end{gather}
(see also \cite{S99,S00,S02}).
The compatibility condition of the system
(\ref{E:Mlin},\ref{E:Ldlin}), which is of course a natural consequence
of the compatibility of Eqs.~\eqref{E:eig1} and \eqref{E:eigtra},
takes the form:
\begin{equation}
A_j \left( M_j L_j^d - L_j^d M_j \right) + 2 A_j^\prime L_j^d = 0,
\label{E:Mldcc}
\end{equation}
and is satisfied iff \eqref{E:chaing} holds.  This linear problem will
play an important role later on when we derive explicit forms
for the Lax pairs for various reductions of the dressing chains.

\subsection{The specialization that contains $P_{III}-P_{VI}$}\label{S:spec}
The above is as far as we will go with the general setting. Now we
specialize to the $\lambda$-dependence \cite{A94}
\begin{equation}
u_j(\lambda,x)=-\lambda^2+\lambda v_j(x)+w_j(x)\label{E:Pgenpot}
\end{equation}
for the potential, and
\begin{equation}
F_j(\lambda,x)=\lambda h_j(x)+f_j(x)\label{E:PgenF}
\end{equation}
for the Darboux transformation. We will now show that this simple case
is still rich enough to allow for three different infinite hierarchies
of equations, containing Painlev\'e equations as their simplest
nontrivial members.

With the above assumptions we find from \eqref{E:Ftil} and
\eqref{E:gF} that both $\left(\log A_j(\lambda,x)\right)'$ and
$A_j(\lambda,x)^{-2}$ should be polynomials in $\lambda$. This is only
possible if the $\lambda$-dependence in $A_j(\lambda,x)$ is
multiplicative, in which case it can always be incorporated into the
integration constants $\mu_j(\lambda)$ in \eqref{E:gF}. Furthermore,
comparing \eqref{E:gF} and \eqref{E:gFd} shows that $\mu_j(\lambda)$
should contain an overall factor of $\lambda-\nu_j$ and thus we can
take it to have the form
\begin{equation}
\mu(j,\lambda)=(\lambda-\nu_j)(\lambda \alpha_j+\beta_j).\label{E:Pgenmu}
\end{equation}

When the above ansatze are substituted into equation \eqref{E:gF} we
find, at successive powers of $\lambda$, the following equations:
\begin{eqnarray}
\lambda^2 : &&\qquad A_j(x)^{-2}\alpha_j=h_j(x)^2 - 1,\label{E:la2}\\
\lambda : &&\qquad A_j(x)^{-2}(\beta_j - \alpha_j\nu_j)=
h_j(x)'+2f_j(x)h_j(x)+v_j(x),\label{E:la1}\\
1 : &&\quad -A_j(x)^{-2}\beta_j\nu_j=f_j(x)'+f_j(x)^2+w_j(x).\label{E:la0}
\end{eqnarray}
The solutions of this set of equations naturally split into 3 distinct
branches:
\begin{enumerate}
\item If $h_j^2\equiv 1$ then we can take (without loss of generality)
  $h_j=1,\,\alpha_j$=0 and $\beta_j= 2$, and solve
  $A_j^{-2}=f_j(x)+\tfrac12 v_j(x)$ from \eqref{E:la1}. The chain
  equations \eqref{E:chaing} obtained from this branch will contain
  $P_{III}$ as a special case.
\item If $h_j\equiv 0$ we can take $\alpha_j= -1, A_j(x)\equiv1$ and
  equation \eqref{E:la1} then becomes $v_j(x)=\nu_j+\beta_j$. For
  convenience we set $v_j(x)\equiv0$ and $\beta_j= -\nu_j$, which
  implies that $\mu(j,\lambda)= \nu_j^2 - \lambda^2$ and
  $u_j(\lambda,x)=-\lambda^2+w_j$ and we therefore obtain the usual
  dressing chain for the Schr\"odinger equation \cite{A94,AS00}, after
  changing notation $\lambda^2\to\lambda,\,\nu_j^2\to\nu_j$. This
  branch contains $P_{IV}$ and $P_V$.
\item Finally, in the generic case we choose $\alpha_j = -1$ so as to
  get $A_j(x)^{-2}=1-h_j(x)^2$.  Equations \eqref{E:la1} and
  \eqref{E:la0} then yield ODEs for $h$ and $f$. In Part 2, this
  branch will be shown to contain $P_{VI}$.
\end{enumerate}
Note also that the $P_{IV}$,$P_V$ branch is obtained as a limit $h\to
0$ from the $P_{VI}$ branch.  The $P_{III}$ branch can also be
obtained as a limit from the $P_{VI}$ branch, but the limit is
singular: Let
\begin{equation}
h_j = 1 - \eps(f_j + \tfrac12 v_j),\quad \alpha_j = -2 \eps,
\label{E:singlim}
\end{equation}
and then as $\eps\to 0$ the leading terms from equations
(\ref{E:la2},\ref{E:la1}) yield $A_j^{-2} = f_j + \tfrac12 v_j$ and $\beta_j
= 2$, respectively.

In all these branches the Painlev\'e equations are obtained when the
derived chain equations are closed (with ``more or less'' periodic
boundary conditions as we shall see later on) after 2, 3 or 4 steps.
Each time there exists a ``hierarchy'' of equations, obtained by
closing the chain after a higher number of steps, but we will not
discuss these higher order equations in the present paper.

\section{The $\pmb{P_{IV-V}}$ branch in general}
\subsection{The generic chain equations}
The simplest realization of the above scheme is found in the case of the
(ordinary) Schr\"odinger equation $(\partial_x^2 + w(x) - \lambda)
\psi(\lambda,x) = 0$, i.e., for a generalized potential $u_j(\lambda,x)$ in
the
operator \eqref{E:eig1} of the type
\begin{equation}
u_j(\lambda,x) = w_j(x) - \lambda.\label{E:PIVpot}
\end{equation}
In this case it is well known that the usual Darboux scheme works
\cite{MS91} and that the $F_j$ are actually $\lambda$-independent
functions $F_j(\lambda,x)=f_j(x)$. More precisely, as explained above,
we must choose particular eigenfunctions $\fie_j(\nu_j)$ of the
Schr\"odinger operator $\partial_x^2 + w_j(x)$ (with eigenvalues
$\nu_j$) such that the $f_j(x)$ are expressed as
\begin{equation}
f_j(x) := (\log \fie_j)_x,
\label{E:Fdef}
\end{equation}
hence these latter functions have to satisfy the following
specialization of relation \eqref{E:gFd}
\begin{equation} 
f_j(x)^\prime + f_j(x)^2 + w_j(x) - \nu_j = 0.\label{E:PIVUj}
\end{equation}
Simultaneously we have to satisfy equation \eqref{E:gF} and  therefore
we choose without loss of generality,
\begin{equation}
\mu_j(\lambda) = (\nu_j - \lambda),\quad A_j(x) = 1.\label{E:PIVAmu}
\end{equation}
This then yields the standard Darboux transformation for the
Schr\"odinger operator with $\tilde{f}_j \equiv f_j$ (from
\eqref{E:Ftil}), $G_j(x) \equiv (\partial_x - f_j(x)) $ (from
\eqref{E:tra1}), and with a change in the potentials \eqref{E:gu}
given by:
\begin{equation}
w_{j+1} = w_j + 2 f_j^\prime \equiv w_j + 2 \left(\log 
\fie_j\right)_{2 x}.\label{E:PIVUtra}
\end{equation}

The resulting chain equation \eqref{E:chaing} for such Darboux
transformations is the well known {\em dressing chain\ } \cite{S92}
\begin{equation}
f_j^\prime + f_{j+1}^\prime = f_j^2 - f_{j+1}^2 + \alpha_j,
\qquad \alpha_j = \nu_{j+1} - \nu_j.
\label{E:PIVchain}
\end{equation}

The linear problem for the dressing chain \cite{S00} follows from the general
expressions (\ref{E:Mop},\ref{E:Mlin}) and (\ref{E:Ldop},
\ref{E:Ldlin}), subject to \eqref{E:PIVAmu}:
\begin{eqnarray}
\left[\partial_x - \Sop - f_j\right] \psi_j &=& 0 \label{E:PIVlp1}\\
\left[\Sop^2 + f_j\Sop + \Sop f_j + \nu_j - \lambda\right] \psi_j &=& 0.
\label{E:PIVlp2}
\end{eqnarray}
From equation \eqref{E:Mldcc} it can be seen that the compatibility
condition of this linear system takes the form of a straightforward
operator commutation relation $[M_j,L_j^d]=0$.

\subsection{Generic bilinearization}
Note that the initial potential, say $w_0$, in the sequence
\eqref{E:PIVUtra} never appears in the dressing chains. In fact, it
only becomes relevant when actually (re-)interpreting the solutions of
these chains -- through \eqref{E:Fdef} -- as solutions of
Schr\"odinger equations such as \eqref{E:PIVUj}, which explicitly
depend on specific potentials. One instance where the underlying
Schr\"odinger equations are of importance is when one wishes to obtain
a Hirota bilinear form for the dressing chain. For, if in the sequence
of coupled Schr\"odinger equations
\begin{equation}
(\partial_x^2 + w_j) \fie_j = \nu_j \fie_j,\, (j=1,2,3),\label{E:p4sch}
\end{equation}
with potentials transforming as in \eqref{E:PIVUtra}, we parameterize
each potential $w_j$ as
\[
w_j = 2 (\log\omega_{j-1})_{2x},
\]
then we obtain a multiplicative transformation rule for the new
functions $\omega_j$
\[
\omega_j=\omega_{j-1}\fie_j.
\]
This suggests a parameterization of the eigenfunctions $\fie_j$
\eqref{E:p4sch} as the ratio of $\omega_j$ and $\omega_{j-1}$.  As is
well known, the standard bilinearization of such Schr\"odinger
equations is through a ratio of $\tau$-functions. Here, with the
benefit of (considerable) hindsight we set
\begin{equation}
\fie_j = \frac{\tau_j}{\tau_{j-1}} e^{-\heps x^2/4},\quad
\omega_j=\tau_j e^{-\heps^2 x^4/96} e^{ - (j-1/2) \heps x^2/4},
\quad \nu_j = \heps (1-j) + \kappa_{j-1},\label{E:PIVbiltra}
\end{equation}
and then \eqref{E:p4sch} transforms into
\begin{equation}
(D_x^2 - \heps x D_x - \kappa_{j-1})\tau_j\cdot \tau_{j-1}=0,
\label{E:bilchaing}
\end{equation}
for some constants $\heps$ and $\kappa_j$. (For a definition of the
Hirota $D$-operators and an introduction to their importance in the
context of the Painlev\'e equations we refer to the review paper
\cite{H99}.)

It can be shown that this chain of bilinear equations is nothing but a
similarity reduction of the $(2+1)$-dimensional dressing chain associated
to the modified KP hierarchy. This entitles us to refer to these
$\tau_j$ as genuine {\em tau-functions} in the sense of Sato theory
\cite{S81,JM83,OS88};  the specific description of this reduction will
be addressed in a separate publication \cite{HW03}.  Note that, as
pointed out above, the initial tau-function $\tau_0$ (and hence the
potential $w_0$) appears explicitly in the bilinear chain
\eqref{E:bilchaing}. 

Below we will see that the system \eqref{E:bilchaing} provides a generic 
bilinearization for all Painlev\'e equations (and higher order variants) 
contained in the dressing chain. There do exist other possibilities when 
it comes to bilinearizing the Painlev\'e equations,
see \cite{HK92,H99,HW03}, but we believe the present generic approach to 
be new. (For a general survey of the Painlev\'e
equations as similarity or symmetry reductions of integrable
nonlinear partial differential equations, see, e.g., \cite{CW99}.)

\subsection{Periodic closing of the dressing chain}
As was mentioned in the introduction we are interested in dressing
chains of finite length. Specifically, we will impose {\em periodic
  closing conditions} on the sequences of Darboux transformations used
in the construction of the chains, i.e., following Adler \cite{AS00},
we require that -- up to a shift in the eigenvalues $\nu_j$ -- the
eigenfunctions $\fie_j(\nu_j)$ which define the Darboux
transformations (as in \eqref{E:Fdef}) become periodic with some
period $N\geq 1$:
\begin{equation}
\fie_{j+N}(\nu_{j+N}) = \fie_j(\nu_j)\qquad\quad\text{with}
\qquad \nu_{j+N} = \nu_j - \eps,\label{E:PIVfieper}
\end{equation}
for a shift $\eps \neq 0$.

As these functions satisfy $(\partial_x^2 + w_j) \fie_j = \nu_j \fie_j$, we
immediately find that the periodic closing of the sequence of functions
$\fie_j$ also implies that the sequence of potentials $w_j$ (generated
through these very Darboux transformations) has to close as 
\begin{equation}
w_{j+N}(x) \equiv w_j(x) - \eps,
\end{equation}
with the obvious implication that the sequence of generalized
potentials $u_j(\lambda,x)$ \eqref{E:PIVpot} closes as
\begin{equation}
u_{j+N}(\lambda,x) \equiv u_j(\lambda + \eps,x),\label{E:PIVgpotper}
\end{equation}
for generic eigenvalues $\lambda$.

The closing conditions \eqref{E:PIVfieper} take on a particularly
simple form when expressed on the bilinear chain \eqref{E:bilchaing}:
\begin{equation}
\tau_{j+N} = \tau_j,\qquad \kappa_{j+N} = \kappa_j,\qquad\text{and}
\quad\heps = \frac{\eps}{N}.\label{E:PIVtaucon}
\end{equation}
It should be remarked that as a consequence one also obtains the constraint
\begin{equation}
\prod_{j = 1}^{N} \fie_j = e^{-\eps x^2/4},
\end{equation}
from \eqref{E:PIVbiltra}.
Finally, for the variables (and parameters) which appear in the
dressing chains \eqref{E:PIVchain}, the periodicity conditions take
the form:
\begin{equation}
f_{j+N} = f_j,\qquad \alpha_{j+N} = \alpha_j,\qquad 
\sum_{j=1}^{N} \alpha_j = -\eps,\label{E:PIVfalfcon}
\end{equation}
accompanied by the first integral (which holds for all $N > 0$)
\begin{equation}
\sum_{j=1}^{N} f_j = -\frac{\eps x}{2}.\label{E:PIVconlaw}
\end{equation}
If $N$ is even there is another integral, obtained by summing every other term
of \eqref{E:PIVchain}:
\begin{equation}
\sum_{j=1}^{N} (-1)^{j}f_j^2 = -\tfrac12\sum_{j=1}^{N} (-1)^{j}\alpha_j
\label{E:PIVconlaw2}
\end{equation}
Note that the cases $N=1,2$ are therefore solvable by quadratures.

\section{$\pmb{P_{IV}}$}
As was discovered almost a decade ago \cite{VS93,A94}, the 3-periodic
dressing chain \eqref{E:PIVchain} is nothing but the $P_{IV}$
equation. We shall now go on to show this explicitly on the dressing
chain, after which we shall derive a bilinear representation and a Lax
pair for the $P_{IV}$-equation.

\subsection{The symmetric form of $P_{IV}$}
Using condition \eqref{E:PIVfalfcon} at $N=3$ we obtain from
\eqref{E:PIVchain} the chain of equations
\begin{equation}
\left\{\begin{array}{l}
f_1' + f_2' = f_1^2 - f_2^2 + \alpha_1,\\
f_2' + f_3' = f_2^2 - f_3^2 + \alpha_2,\\
f_3' + f_1' = f_3^2 - f_1^2 + \alpha_3,\\
\end{array}\right. \label{E:PIVchain3}
\end{equation}
with
\begin{equation}
\alpha_1+\alpha_2+\alpha_3 = -\eps\label{E:PIValphacon}.
\end{equation}
Expressed in terms of
\begin{equation}
\begin{array}{l}
g_1 = f_1 + f_2 = (\log \fie_1 \fie_2)_x,\\
g_2 = f_2 + f_3 = (\log \fie_2 \fie_3)_x,\\
g_3 = f_3+ f_1 = (\log \fie_3 \fie_1)_x,
\end{array}\label{E:PIVgdef}
\end{equation}
we get
\begin{gather}
\left\{\begin{array}{l}
g^\prime_1 = g_1(g_3-g_2) + \alpha_1,\\
g^\prime_2 = g_2(g_1-g_3) + \alpha_2,\\
g^\prime_3 = g_3(g_2-g_1) + \alpha_3. 
\end{array}\right.\label{E:PIVsym}
\end{gather}
Following Noumi and Yamada \cite{NY99,N00} we shall refer to this
system as the {\em symmetric form} of the $P_{IV}$ equation.  This
particular form of the period 3 dressing chain appears already in
\cite{A94}. However, it is worth pointing out that \eqref{E:PIVsym} is
already presented in \cite{Bu} where it is used to integrate a higher
order non linear differential equation in terms of the $P_{IV}$
equation.

Because of the constraint \eqref {E:PIValphacon} -- or alternatively,
as a consequence of \eqref{E:PIVconlaw} -- the system \eqref{E:PIVsym}
can be integrated once:
\begin{equation}
g_1 + g_2 + g_3 = -\eps x\label{E:PIVgcon}
\end{equation}
(where a possible integration constant only amounts to a translation in
$x$ and can therefore be omitted).
Eliminating $g_3$ we get
\begin{equation}
\left\{\begin{array}{l}
g_1^\prime = \alpha_1 - 2 g_1 g_2 - \eps x g_1 - g_1^2,\\
g_2^\prime = \alpha_2 + 2 g_1 g_2 + \eps x g_2 + g_2^2.
\end{array}\right.\label{E:pIVpair}
\end{equation}
These equations are in Hamiltonian form:
\begin{equation}
  g_1' = \frac{\partial H}{\partial g_2},
\qquad g_2' = -\frac{\partial H}{\partial g_1},\label{E:Ham}
\end{equation}
where
\begin{equation}
 H = -g_1 g_2^2 - g_2 g_1^2 - \eps x g_1 g_2 - \alpha_2 g_1
  + \alpha_1 g_2.
\end{equation}
Note that this Hamiltonian differs slightly (by a canonical
transformation) from the one discussed by Okamoto \cite{O81,O86a,O99};
it does, however, appear in this form in the classic work \cite{M23},
and in \cite{NY99}.

The $P_{IV}$ equation is now obtained if we eliminate $g_2$ and denote
$y(z)=\kappa g_1(x)$, where $x=\kappa z, \,\kappa^2=2/\eps$, this
yields the standard form
\begin{equation}
\frac{d^2 y}{dz^2} = \frac{1}{2 y} \left(\frac{d y}{d z}\right)^2 + 
\frac{3}{2} y^3 + 4 z y^2 + 2 ( z^2 - a) y + \frac{b}{y}
\label{E:PIV}
\end{equation}
where the parameters $a$ and $b$ are given by
\begin{equation}
a = (\alpha_2 -\alpha_3)/\eps,\qquad
b =-2(\alpha_1/\eps)^2.
%a = \frac{\alpha_3 -\alpha_2}{\alpha_1+\alpha_2+\alpha_3},\qquad\qquad 
%b =-2\left(\frac{\alpha_1}{\alpha_1+\alpha_2+\alpha_3}\right)^2.
\end{equation}

\subsection{Bilinear form of $P_{IV}$}
Imposing conditions \eqref{E:PIVtaucon} on the bilinear form
\eqref{E:bilchaing} at $N=3$ we immediately obtain a bilinearization of the
$P_{IV}$ equation:
\begin{equation}
\left\{\begin{array}{l}
(D_x^2  - \frac{\eps x}{3} D_x - \kappa_0) \tau_1 \cdot \tau_0 = 0,\\
(D_x^2  - \frac{\eps x}{3} D_x - \kappa_1) \tau_2 \cdot \tau_1 = 0,\\
(D_x^2  - \frac{\eps x}{3} D_x - \kappa_2) \tau_0 \cdot \tau_2 = 0.
\end{array}\right.\label{E:PIVbilfo}
\end{equation}
The transformation for the $g_i$ is:
\begin{equation}
g_i = \partial_x\log \frac{\tau_{i+1}}{\tau_{i-1}}
 - \frac{\eps x}{3} ,
\qquad \alpha_i\equiv \kappa_i - \kappa_{i-1} - \frac{\eps}{3},
\label{E:PIVbiltra1}
\end{equation}
for $i=1,2,3$, and with periodicity \eqref{E:PIVtaucon} (see also
\cite{NY99,N00}).  Recall that \eqref{E:PIVbilfo} was obtained by making
the substitution \eqref{E:PIVbiltra} directly into the (now periodic) chain of
Schr\"odinger equations \eqref{E:p4sch}.  If, on the other hand, we make
the substitutions \eqref{E:PIVbiltra1} into the
\hyphenation{sym-me-tric}symmetric form \eqref{E:PIVsym} we will only
obtain two equations for the three tau-functions.  This is because the
equations \eqref{E:PIVsym} do not contain any information on the
potentials $w$. However, if we explicitly include the assumption $w_j
= 2 (\log\tau_{j-1})_{2x}$, e.g., in the form of the first equation of
\eqref{E:PIVbilfo}, then we do obtain a well-determined system of
bilinear equations. Note also that
\begin{equation}
y(z)=\partial_z\log  \left(\frac{\tau_{2}}{\tau_{0}}\,
e^{-\frac{z^2}{3}}\right).\label{ytau4}
\end{equation}

\subsection{Lax pair for $P_{IV}$}\label{S4.3}
Let us see what the consequences of the periodic closing \eqref{E:PIVfieper}
are on the linear problem (\ref{E:PIVlp1},\ref{E:PIVlp2}). According to
\eqref{E:PIVgpotper} the generalized potentials $u_j(\lambda,x)$ are periodic
for generic $\lambda$ as well (up to a shift in $\lambda$) and thus we may
require similar periodicity for some eigenfunctions of the linear problem
\eqref{E:eig1}, namely those generated by successive Darboux transformations:
\begin{equation}
\psi_{j+3}(\lambda,x) = \psi_j(\lambda + \eps,x).
\end{equation}
The linear system (\ref{E:PIVlp1},\ref{E:PIVlp2}) therefore reduces to 
the following set of ``difference equations'' in the spectral parameter
$\lambda$:
\begin{equation}
\partial_x\begin{pmatrix}
\psi_1\\\psi_2\\\psi_3
\end{pmatrix} = \begin{pmatrix} \psi_2 + f_1 \psi_1 \\ \psi_3 + f_2 \psi_2
\\\psi_1^s + f_3 \psi_3 \end{pmatrix} ,\qquad
 \begin{pmatrix} \psi_3 + g_1 \psi_2 + \nu_1 \psi_1 \\
 \psi_1^s + g_2 \psi_3 + \nu_2 \psi_2
 \\\psi_2^s + g_3 {\psi}_1^s +
 \nu_3 \psi_3 \end{pmatrix} = \lambda \begin{pmatrix}
 \psi_1\\\psi_2\\\psi_3\end{pmatrix},\label{E:PIVdellp}
\end{equation}
where the symbol $\psi^s_j$ denotes eigenfunctions at shifted values
of the spectral parameter: $\psi^s_j(\lambda,x) := \psi_j(\lambda+\eps,x)$.

In order to recover a more customary linear problem for the $P_{IV}$
equation, let us first rewrite the equations \eqref{E:PIVdellp} as:
\begin{gather}
  \partial_x\Psi(\lambda) = {\cal B}_1 \Psi(\lambda) + {\cal
    B}_2 \Psi(\lambda + \eps),\label{E:PIVdellp1}\\
  \left({\cal A}_1 - \lambda \mathbb{I}\right) \Psi(\lambda) + {\cal A}_2
  \Psi(\lambda+\eps) = 0,
\end{gather}
for $\Psi(\lambda) =
(\psi_1(\lambda,x),\psi_2(\lambda,x),\psi_3(\lambda,x))^t$ and with
matrices
${\cal B}_1, {\cal B}_2, {\cal A}_1$ and ${\cal A}_2$
\begin{equation}
{\cal B}_1 = \begin{pmatrix}f_1&1&0\\0&f_2&1\\0&0&f_3\end{pmatrix},\ {\cal
B}_2 =  \begin{pmatrix}0&0&0\\0&0&0\\1&0&0\end{pmatrix},\ {\cal A}_1 =
\begin{pmatrix}\nu_1&g_1&1\\0&\nu_2&g_2\\0&0&\nu_3\end{pmatrix},\
{\cal A}_2 = \begin{pmatrix}0&0&0\\1&0&0\\g_3&1&0\end{pmatrix}.
\end{equation}
If we now define the (formal) Fourier transform of $\Psi(\lambda)$ as
\begin{equation}
\tilde{\Phi}(k) \equiv \int\ {\rm d}\lambda\ e^{i k
\lambda}\ \Psi(\lambda) ,\label{E:PIVfourier}
\end{equation}
we immediately see that $\tilde{\Phi}(k)$ satisfies local
relations in $k$, for example: 
\begin{equation}
\partial_x\tilde{\Phi}(k) = {\cal B}_1 \tilde{\Phi}(k) + {\cal B}_2 e^{-i \eps
k} \tilde{\Phi}(k)
\end{equation}
instead of the original difference relation \eqref{E:PIVdellp1} in
$\lambda$.
Consequently, introducing a new parameter $\xi$ and a new function
$\Phi(\xi)$
in terms of $\tilde{\Phi}(k)$ by
\begin{equation}
\xi: = e^{-i \eps k} ,\qquad \Phi(\xi) := \left.\tilde{\Phi}(k)
\right|_{k=\frac{i}{\eps} \log\xi} ,\label{E:Mellin}
\end{equation}
we find that $\Phi(\xi)$ satisfies the following Lax pair for the
$P_{IV}$ equation,
\begin{equation}
\left\{\begin{array}{r}
\partial_x\Phi = {\cal M} \Phi ,\qquad {\cal M} \equiv {\cal B}_1 + 
\xi {\cal B}_2,\\
-\eps \xi \partial_{\xi}\Phi = {\cal L} \Phi ,\qquad {\cal L} = 
{\cal A}_1 + \xi {\cal A}_2,
\end{array}\right.
\end{equation}
and the compatibility condition $\partial_x{\cal L} + \eps \xi
\partial_{\xi}{\cal M} = [{\cal M}, {\cal L}]_-$ is nothing but the
symmetric form \eqref{E:PIVsym} for $P_{IV}$. This Lax pair already
appears in \cite{N00}, in connection with reductions of the
Drinfeld-Sokolov hierarchy. Also, the transformations
\eqref{E:PIVfourier} and \eqref{E:Mellin} combined, amount to the
Mellin transformation that is used (to the same effect) in \cite{T02}
in order to connect Lax pairs arising in the context of dressing
chains to those that appear in the work by Noumi et al.

\section{$\pmb{P_V}$}
Similarly to the above, $P_V$ can be obtained by closing the dressing
chain \eqref{E:PIVchain} at $N=4$ \cite{VS93,A94}.  Just as for the
$P_{IV}$ equation this insight yields immediate access to a Lax pair
and bilinear formulation for the $P_V$ equation.

\subsection{Periodic closing and $P_V$}
Closing the Darboux chain \eqref{E:PIVchain} at period $N=4$ 
(cf. condition \eqref{E:PIVfalfcon}) yields the following system of
differential equations:
\begin{equation}
\left\{\begin{array}{l}
f_1' + f_2' = f_1^2 - f_2^2 + \alpha_1,\\
f_2' + f_3' = f_2^2 - f_3^2 + \alpha_2,\\
f_3' + f_4' = f_3^2 - f_4^2 + \alpha_3,\\
f_4' + f_1' = f_4^2 - f_1^2 + \alpha_4,
\end{array}\right. \label{E:PVchain}
\end{equation}
with constants $\alpha_j$ ($j=1,2,3,4$) subject to the
constraint
\begin{equation}
\alpha_1 + \alpha_2 + \alpha_3 + \alpha_4 = -\eps.
\end{equation}
Now we have two first integrals
\begin{eqnarray}
f_1 + f_2 + f_3 + f_4 &=& -\tfrac12{\eps} x,\label{E:PVI1} \\
- f_1^2 + f_2^2 - f_3^2 + f_4^2 &=& 
\tfrac12({\alpha_1  - \alpha_2 + \alpha_3 - \alpha_4}) =: \omega,
\label{E:PVI2}
\end{eqnarray}
(recall (\ref{E:PIVconlaw},\ref{E:PIVconlaw2})).  Hence, the system
\eqref{E:PVchain} can be reduced to a $2^{\rm{nd}}$ order differential
equation, which turns out to be nothing but the Painlev\'e $V$ equation. We
shall now proceed to show this.

It is convenient to rewrite the symmetric form \eqref{E:PVchain} in
terms of new variables $g_1, g_2$ obtained by resolving \eqref{E:PVI1}
and \eqref{E:PVI2} as
\begin{eqnarray*} 
f_1&=&-\tfrac1{2\eps x} g_1 g_2 +\tfrac14 (g_1 - g_2)
 - \tfrac{\eps x }8 + \tfrac{\omega}{\eps x},\\ 
f_2&=&\phantom{-}\tfrac1{2\eps x} g_1 g_2 +\tfrac14 (g_1 + g_2)
 - \tfrac{\eps x }8 - \tfrac{\omega}{\eps x},\\ 
f_3&=&-\tfrac1{2\eps x} g_1 g_2 -\tfrac14 (g_1 - g_2)
 - \tfrac{\eps x }8 + \tfrac{\omega}{\eps x},\\ 
f_4&=&\phantom{-}\tfrac1{2\eps x} g_1 g_2 -\tfrac14 (g_1 + g_2)
 - \tfrac{\eps x }8 - \tfrac{\omega}{\eps x},
\end{eqnarray*} 
the inverse relation of which is
\begin{eqnarray*}
g_1&=&\phantom{-}f_1+f_2-f_3-f_4,\\
g_2&=&-f_1+f_2+f_3-f_4,
\end{eqnarray*} 
together with the constraints \eqref{E:PVI1} and \eqref{E:PVI2}.
In terms of these new free functions $g_1$ and $g_2$  we obtain
the symmetric form
\begin{eqnarray} 
g_1'&=&-\tfrac1{\eps x} g_1^2 g_2+\tfrac{2\omega}{\eps x} g_1
 + \tfrac{\eps x}4 g_2 +\alpha_1-\alpha_3,\label{E:p5g1e}\\ 
g_2'&=&\phantom{-} \tfrac1{\eps x} g_1 g_2^2-\tfrac{2\omega}{\eps x} g_2
 - \tfrac{\eps x}4 g_1 +\alpha_2-\alpha_4,\label{E:p5g2e} 
\end{eqnarray} 
which is Hamiltonian \eqref{E:Ham} with
\begin{equation}
H:=-\tfrac1{2\eps x} g_1^2 g_2^2+\tfrac{\eps x}8(g_1^2+g_2^2)
+\tfrac{2\omega}{\eps x}g_1g_2 -(\alpha_2 -\alpha_4)g_1
+(\alpha_1-\alpha_3)g_2,
\end{equation}
Note that this Hamiltonian differs from the form used in
\cite{M23,O81,O87b,O99}, but they are connected by a canonical
transformation and a change of the independent variable.

The $P_V$ equation is now obtained if we first solve $g_2$ from
\eqref{E:p5g1e}, substitute it into \eqref{E:p5g2e} and express
$g_1$ in terms of  $y$ defined by 
\[ 
y=\tfrac12-\tfrac1{\varepsilon x}\,g_1,
\quad \text{where}\quad y=y(z),\,z=\eps x^2/4,
\] 
this yields the $P_V$ equation in the form (used, e.g., in
\cite{HK92,H99}, up to an extra transformation of the independent
variable)
\begin{align}
\frac{d^2 y}{d z^2} = &\left(\frac{1}{2 y} + \frac{1}{2(y-1)}\right)
\left(\frac{dy}{dz}\right)^2-\frac1z\frac{dy}{dz}\nonumber\\
&-\frac{\alpha y}{z^2(y-1)} - \frac{\beta(y-1)}{z^2y}
-\frac{\gamma y(y-1)}{z}-\delta y(y-1)(2y-1),
\end{align}
where
\[
\alpha=\frac{(\alpha_1-\alpha_3-\omega+\tfrac12\varepsilon)^2}{8\varepsilon^2},
\quad
\beta=-\frac{(\alpha_1-\alpha_3+\omega-\tfrac12\varepsilon)^2}{8\varepsilon^2},
\quad \gamma=\frac{\alpha_4-\alpha_2}{\varepsilon},
\quad \delta=-\frac1{2}.
\]
The usual form of $P_V$ is obtained for $w=y/(y-1)$, which permutes
the poles at 1 and $\infty$.

\subsection{Bilinear form of $P_V$}
Imposing conditions \eqref{E:PIVtaucon} on the bilinear form
\eqref{E:bilchaing} of the dressing chain, this time at $N=4$, we obtain a
bilinearization of the $P_{V}$ equation in terms of 4 tau-functions:
\begin{equation}
\left\{\begin{array}{l}
(D_x^2  - \frac{\eps x}{4} D_x - \kappa_0) \tau_1 \cdot \tau_0 = 0\\
(D_x^2  - \frac{\eps x}{4} D_x - \kappa_1) \tau_2 \cdot \tau_1 = 0\\
(D_x^2  - \frac{\eps x}{4} D_x - \kappa_2) \tau_3 \cdot \tau_2 = 0\\
(D_x^2  - \frac{\eps x}{4} D_x - \kappa_3) \tau_0 \cdot \tau_3 = 0
\end{array}\right..\label{E:PVbilfo}
\end{equation}
Subject to the same remarks as in the case of the $P_{IV}$ equation, this
system of Hirota equations gives a bilinear representation for the symmetric
form \eqref{E:PVchain} using substitution
\begin{equation}
f_i = \partial_x\left(\log \frac{\tau_i}{\tau_{i-1}}\right) - 
\frac{\eps x}{8} ,
\qquad \alpha_i\equiv \kappa_i - \kappa_{i-1} - \frac{\eps}{4},
\end{equation}
for $i=1,2,3,4$, and with periodicity \eqref{E:PIVtaucon}.
Note also that (cf. \eqref{ytau4})
\begin{equation}
y(z)=\partial_z\log\left(\frac{\tau_0}{\tau_2} 
e^{\frac{z}{2}}\right).\label{ytau5}
\end{equation}

It was already mentioned that in \cite{NY98a} Noumi and Yamada
presented a class of dynamical systems, each member of which possesses
a particular symmetry group of (affine) Weyl-type
$\widehat{W}(A_{n-1}^{(1)})$. $P_{IV}$ and $P_V$ are contained in this
class at levels $n=3$ and $4$ respectively. In can be shown
\cite{W03,HW03}, not only that the bilinear forms of the $P_{IV}$ and
$P_V$ equations (\eqref{E:PIVbilfo} and \eqref{E:PVbilfo}) possess
similar symmetries, but generally that {\em any} bilinear system
contained in \eqref{E:bilchaing} possesses a symmetry group of type
$\widehat{W}(A_{n-1}^{(1)})$. Hence these equations are nothing but
the bilinear formulations of the Noumi-Yamada systems and
consequently, we have shown that each member in that class corresponds
to a periodic dressing chain (in the sense of Sects. 3.2 and 3.3).

\subsection{Lax pair for $P_V$}
Exactly as in the $P_{IV}$ case one can obtain a Lax pair for $P_V$ by
imposing periodicity on the generic Lax pair
(\ref{E:PIVlp1},\ref{E:PIVlp2}) for the dressing chain:
\begin{equation}
\psi_{j+4}(\lambda,x) = \psi_j(\lambda+\eps,x).
\end{equation}
Following the notation of Sec.~\ref{S4.3} we can write:
\begin{equation}
\partial_x \begin{pmatrix}\psi_1\\\psi_2\\\psi_3\\\psi_4\end{pmatrix} = 
\begin{pmatrix}\psi_2 + f_1 \psi_1\\\psi_3 + f_2 \psi_2\\\psi_4 + f_3 
\psi_3\\\psi_1^s + f_4 \psi_4\end{pmatrix},
\qquad
\begin{pmatrix}\psi_3 + (f_1 + f_2) \psi_2 + \nu_1 \psi_1\\\psi_4 + 
(f_2 + f_3) \psi_3 + \nu_2 \psi_2\\\psi_1^s + (f_3 + f_4) \psi_4 + 
\nu_3 \psi_3\\\psi_2^s + (f_4 + f_1) \psi_1^s + \nu_4 \psi_4
\end{pmatrix} = \lambda \begin{pmatrix}\psi_1\\\psi_2\\\psi_3\\\psi_4
\end{pmatrix}.
\end{equation}
After a (formal) Fourier transformation similar to the one in the
$P_{IV}$ case (\ref{E:PIVfourier},\ref{E:Mellin}), we obtain the
following Lax formulation for the $P_V$ equation \cite{N00}:
\begin{equation}
\partial_x\Phi = {\cal M} \Phi, \qquad -\eps \xi \partial_{\xi}\Phi
 = {\cal L} \Phi,
\end{equation}
\begin{equation}
{\cal M} = \begin{pmatrix}f_1 & 1 & 0 & 0\\ 0 & f_2 & 1 & 0\\
 0 & 0 & f_3 & 1\\ 0 & 0 & 0 & f_4\end{pmatrix} + \xi 
\begin{pmatrix}0 & 0 & 0 & 0\\ 0 & 0 & 0 & 0\\0 & 0& 0 & 0\\
 1 & 0 & 0 & 0\end{pmatrix},
\end{equation}
\begin{equation}
{\cal L} = \begin{pmatrix}\nu_1 & f_1 + f_2 & 1 & 0\\
 0 & \nu_2 & f_2 + f_3 & 1\\ 0 & 0 & \nu_3 & f_3 + f_4\\
 0 & 0 & 0 & \nu_4\end{pmatrix} + \xi 
\begin{pmatrix}0 & 0 & 0 & 0\\ 0 & 0 & 0 & 0\\1 & 0& 0 & 0\\
 f_1 + f_4 & 1 & 0 & 0\end{pmatrix}.
\end{equation}
The compatibility condition $\partial_x {\cal L} + \eps \xi
\partial_{\xi}{\cal M} = [{\cal M}, {\cal L}]_-$ is easily seen to
correspond to the symmetric form \eqref{E:PVchain} of the $P_{V}$
equation.

\section{The $\pmb{P_{III}}$ branch in general}
As was noted in the introduction, the $P_{III}$ branch is quite
different from that for the $P_{IV}-P_{V}$ equations. In this section
we will first study this branch in general and in the next section we
shall close the chain in order to obtain $P_{III}$.

\subsection{The generic chain equations}
As was explained in Section \ref{S:spec}, a different branch of the
chain equation \eqref{E:chaing} for potentials of type
\eqref{E:Pgenpot} is obtained if we choose
\begin{eqnarray}
F_j(\lambda,x)&:=&\lambda+f_j(x),\label{E:PIIIF}\\
\mu_j(\lambda)&:=&2(\lambda-\nu_j),\\
A_j(x)&:=&\left[f_j(x)+ \tfrac12 v_j(x)\right]^{-\frac{1}{2}},\label{E:Adef}
\end{eqnarray}
($f_j+\tfrac12 v_j\not\equiv 0$); the only remaining part of condition
\eqref{E:gF} is the following constraint on $f$ \eqref{E:la0}:
\begin{equation}
f_j'+f_j^2+\nu_j(2f_j+v_j)+w_j=0.
\label{E:Fn}
\end{equation}
For the present transformation to be a genuine Darboux transformation,
we must also require that the intertwiner $G_j$ in \eqref{E:tra1}
annihilates some particular eigenfunction $\fie_j:=\psi(\nu_j,x)$ of
the original Schr\"odinger equation, i.e., $F_j(\nu_j,x) = (\log
\fie_j)_x$ or in terms of $f$:
\begin{equation}
f_j:=-\nu_j+\fie'_j/\fie_j.\label{E:PIIIf}
\end{equation}
Condition \eqref{E:Fn} then turns into the Schr\"odinger equation for the
eigenfunction $\fie_j$:
\begin{equation}
\fie_j''+(-\nu_j^2+\nu_j \,v_j+w_j)\fie_j=0.\label{E:fieP3}
\end{equation}
The transformation induced for the potentials is obtained from \eqref{E:gu}:
\begin{eqnarray}
v_{j+1}&=&v_j+2(\log A_j)',\label{E:v3tr}\\
w_{j+1}&=&w_j+[2f_jA_j-A_j']'/A_j.\label{E:w3tr}
\end{eqnarray}

From the explicit form \eqref{E:Adef} of $A$ it is clear that in the
present case the chain equations generated by \eqref{E:chaing} --
subject to condition \eqref{E:Fn} -- will take the form of a dynamical
system in the variables $f$ and $v$. However, these variables are
strongly coupled and from (\ref{E:Fn},\ref{E:v3tr},\ref{E:w3tr})
we get (after eliminating $w$)
\begin{align}
f_j'+\tfrac12 v_j'&+(f_j+\tfrac12 v_j)(v_{j+1}-v_j)=0,\\
f_j'-\tfrac12 v_j'&
-(f_{j-1}+\nu_{j-1})^2+(\tfrac12 v_{j-1}-\nu_{j-1})^2
+(f_{j}+\nu_{j})^2 -(\tfrac12 v_{j}-\nu_{j})^2=0.
\end{align}
It is convenient to change to new dependent variables $d_j,\,r_j$ defined
by 
\begin{equation}
f_j=\tfrac12 d_j+\tfrac12r_j-\nu_j,\quad v_j=d_j-r_j+2\nu_j,
\quad \beta_j=\nu_j-\nu_{j+1},
\end{equation}
(note that $d_j=f_j+\tfrac12 v_j \equiv A_j^{-2}$) because then the
{\em chain equations of the $P_{III}$ branch} take the following
simple form (which already appears in \cite{S99})
\begin{eqnarray}
d_j{}'&=&d_j(d_j-r_j-d_{j+1}+r_{j+1}+2\beta_j),\label{E:n3c1}\\
r_j{}'&=&d_{j-1}r_{j-1}-d_jr_j.\label{E:n3c2}
\end{eqnarray}

The linear problem for this integrable system is of course obtained
from (\ref{E:Mop},\ref{E:Mlin}) and (\ref{E:Ldop},\ref{E:Ldlin}) by
appropriate identifications of the variables appearing in those
equations, but we shall not discuss its general form here.

\subsection{Generic bilinearization}
It is of interest to bilinearize the chain equations
(\ref{E:n3c1},\ref{E:n3c2}) before closing the chain.  We start by
representing $A$ in terms of a new function $\tilde\fie_j$ and the
eigenfunction $\fie_j$:
\begin{equation}
  \label{E:afie}
  A_j^{-2} =\frac{\tilde\fie_j}{\fie_j},
\end{equation}
which due to \eqref{E:PIIIf} amounts to
\begin{equation}
\fie_j' = \tilde\fie_j + (\nu_j - \tfrac12 v_j) \fie_j.\label{E:feq}
\end{equation}
Since $\fie_j$ is a solution of \eqref{E:fieP3} we also obtain
\begin{equation}
\tilde\fie_j' = \left(\tfrac12 {v_j} -\nu_j\right)  \tilde\fie_j + 
(\tfrac12 {v_j'} - w_j - \tfrac14 {v_j^2}) \fie_j.\label{E:ftileq}
\end{equation}
Taken together, equations (\ref{E:feq},\ref{E:ftileq}) yield a first
order representation of the Schr\"odinger equation \eqref{E:fieP3}
and as such provide a suitable starting point for the bilinearization of the
Schr\"odinger equations associated with the $P_{III}$ branch.

Under the above ansatz, the transformation formula \eqref{E:v3tr}
suggests a logarithmic parameterization for the potential $v$
\begin{equation}
v_j := (\log q_j)',\label{E:PIIIv}
\end{equation}
because then the transformation for $q$ takes the simple multiplicative form
\begin{equation}
q_{j+1} \sim q_j \fie_j/\tilde\fie_j,\label{E:qtra}
\end{equation}
where $\sim$ denotes equivalence up to a constant multiple.  The
transformation formula for $w$ is best expressed in terms of a new variable
$W$
\begin{equation}
w_j = \tfrac12 W_j - \tfrac14 v_j^2,\label{E:Wdef}
\end{equation}
because then we get
\begin{equation}
W_{j+1} = W_j + (\log \fie_j)'' - (\log q_j)'' + (\log \tilde\fie_j)'',
\end{equation}
(from \eqref{E:w3tr} using also (\ref{E:feq},\ref{E:qtra})) which
suggests a parameterization
\begin{equation}
W_j := (\log \omega_j)''. 
\end{equation}
The transformation for $W$ then takes the simple multiplicative form
\begin{equation}
\omega_{j+1} \approx \frac{\omega_j \fie_j \tilde\fie_j}{q_j} \sim 
\frac{\omega_j \fie_j^2}{q_{j+1}}\label{E:omegatra}
\end{equation}
(where $\approx$ denotes equivalence up to a gauge factor $e^{ax+b}$).

If, as before, we think of the functions $\fie_j$, $\tilde\fie_j$,
$q_j$ and $\omega_j$ as ratios of ``tau-functions'', it becomes clear
that
\begin{align}
q_j& = \tau_j^+/\tau_j\ e^{\gamma_j x},
\hspace{2.9cm}
\omega_j = \tau_j \tau_j^+,\label{E:qomtra}\\
\fie_j& = c_j \frac{\tau^+_{j+1}}{\tau^+_{j}}
\sqrt{\frac{\tau_j^+}{\tau_j}}\ e^{\alpha_j x},
\qquad \qquad 
\tilde\fie_j = \tilde{c}_j\ \frac{\tau_{j+1}}{\tau_j} 
\sqrt{\frac{\tau_j^+}{\tau_j}}\ e^{\tilde\alpha_j x},
\label{E:fietilfietra}
\end{align}
is a good representation, since it automatically satisfies the
multiplicative transformation rules for $v$ and $w$, equations
\eqref{E:qtra} and \eqref{E:omegatra}, if the introduced constants
satisfy
\begin{equation}
\gamma_{j+1} = \gamma_j + \alpha_j - \tilde\alpha_j.\label{gamdef}
\end{equation}
For $f$ and $v$ we get from \eqref{E:PIIIf} and \eqref{E:PIIIv}
\begin{equation}
f_j=\tfrac12\partial_x\log\left(\tfrac{{\tau^+_{j+1}}^2}{\tau_j\tau_j^+}
\right)+\alpha_j-\nu_j,\qquad
v_j=\partial_x\log\left(\tfrac{\tau_{j}^+}{\tau_j}\right)+\gamma_j.
\end{equation}
and then
\begin{equation}
d_j=\partial_x\log\left(\tfrac{\tau_{j+1}^+}{\tau_j}\right)+
\alpha_j+\tfrac12\gamma_j-\nu_j,\quad
r_j=\partial_x\log\left(\tfrac{\tau_{j+1}^+}{\tau_j^+}\right)+
\alpha_j-\tfrac12\gamma_j+\nu_j.
\label{E:drbiltrag}
\end{equation}

As mentioned before, the assignments
(\ref{E:qomtra},\ref{E:fietilfietra}) take care of equations
(\ref{E:v3tr},\ref{E:w3tr}), but we still have to study the
Schr\"odinger  equation \eqref{E:fieP3}, which we take in its first
order form (\ref{E:feq},\ref{E:ftileq}). This yields
\begin{equation*}
\left\{\begin{array}{l}
(D_x + \alpha_j - \nu_j + \frac{\gamma_j}{2})\ \tau_{j+1}^+ \cdot \tau_j = 
\frac{\tilde{c}_j}{c_j}\ e^{(\tilde\alpha_j-\alpha_j) x}\ \tau_j^+ 
\tau_{j+1},\\
D_x (D_x + \alpha_j - \nu_j + \frac{\gamma_j}{2})\,
\tau_{j+1}^+ \cdot \tau_j =
\frac{\tilde{c}_j}{c_j}\ e^{(\tilde\alpha_j-\alpha_j) x}\ 
(D_x +\gamma_j -\alpha_j -\tilde\alpha_j - 2\nu_j) \tau_j^+ \cdot \tau_{j+1}.
\end{array}\right.
\end{equation*}
In (\ref{E:qomtra},\ref{E:fietilfietra}) we have some additional
freedom in the parameters $\gamma,\alpha,\tilde\alpha$ (by scaling
$\tau_j^+\to\tau_j^+e^{s_jx},\tau_j\to\tau_je^{-s_jx}$), which can be
used to assign a fixed value to the parameter $\gamma_j$
\begin{equation}
\gamma_j=2(\nu_j-\alpha_j).\label{newgam}
\end{equation}
In order to satisfy \eqref{gamdef} we must then also require that
\begin{equation}
\tilde\alpha_j=2\alpha_{j+1}-\alpha_j-2(\nu_{j+1}-\nu_j).\label{alphatil}
\end{equation}
After this redefinition of parameters, the equations simplify
considerably and we get
\begin{equation}
\left\{\begin{array}{rcl}
D_x \, \tau_{j+1}^+ \cdot \tau_j &=& 
\kappa_j\ e^{\sigma_j x}\,\tau_j^+\tau_{j+1},\\
D_x^2\, \tau_{j+1}^+ \cdot \tau_j &=& 
\kappa_j\ e^{\sigma_j x}\,(D_x -\rho_j) \tau_j^+ \cdot \tau_{j+1}.
\end{array}\right.\label{E:P3bilchaing}
\end{equation}
where
\begin{align*}
  \sigma_j&=\tilde\alpha_j-\alpha_j=
2(\alpha_{j+1}-\alpha_j-\nu_{j+1}+\nu_j),\,\\
  \rho_j&=\tilde\alpha_j+3\alpha_j=2(\alpha_{j+1}+\alpha_j-\nu_{j+1}+\nu_j),\\
  \kappa_j&=\tilde c_j/c_j.
\end{align*}
Equations \eqref{E:P3bilchaing} form the Hirota bilinear representation
of the chain equations (\ref{E:n3c1},\ref{E:n3c2}). For later use note
that due to the first equation of \eqref{E:P3bilchaing} we can also
write
\begin{equation}
d_j=\kappa_j\ e^{\sigma_j x}\frac{\tau_j^+\tau_{j+1}}{\tau_{j+1}^+
  \tau_j}.\label{E:dtprod}
\end{equation}

\subsection{Periodic closing in general}
As before, we are mainly interested in a finite chain of equations
obtained after requiring the potential $u_j(\lambda)$ to be periodic,
up to a shift in the (generic) eigenvalue, as stated in
\eqref{E:PIVgpotper}, and this implies the existence of ``periodic''
eigenfunctions \eqref{E:PIVfieper}.  For the potentials $v,w$ this now
implies
\begin{equation}
v_{j+N} = v_j - 2\eps, \qquad\quad w_{j+N} = w_j + \eps v_j -\eps^2,
\label{vcond}
\end{equation}
whereas the auxiliary potential $W$ of \eqref{E:Wdef} is strictly
periodic: $W_{j+N} = W_j$.

On the level of the $\tau$-functions $\tau, \tau^+$ these closing
conditions translate into strict periodicity
\begin{equation}
\tau_{j+N}^+ = \tau_j^+,\qquad \tau_{j+N} = \tau_j.
\end{equation}
Condition \eqref{gamdef} was replaced by constraints \eqref{newgam}
and \eqref{alphatil}, which are compatible with the requirement that
the parameters $\alpha_j,\,\tilde\alpha_j,\,c_j,\,\tilde
c_j,\,\sigma_j,\, \rho_j,\,\kappa_j$ are strictly periodic and
\begin{equation}
\gamma_{j+N} = \gamma_j - 2 \eps.
\end{equation}
We also have
\begin{equation}
\sum_{j=1}^N \beta_j = \eps,\qquad
\sum_{j=1}^N \sigma_j = 2\eps.\label{E:P3alfcon}
\end{equation}
Furthermore, from (\ref{E:drbiltrag},\ref{newgam}) it is obvious that
$d_j$ and $r_j$ are (strictly) periodic as well and from
\eqref{E:dtprod} and \eqref{E:drbiltrag} we also find the following
two (generic) conservation laws for the periodic chain equations\cite{S99}:
\begin{equation}
\prod_{j=1}^{N} d_j = e^{2\eps x}\,\prod_{j=1}^N\kappa_j ,\qquad
\sum_{j=1}^N r_j=2\,\sum_{j=1}^N\alpha_j.
\label{E:P3conlawg}
\end{equation}

\section{$\pmb{P_{III}}$}
\subsection{Closing the chain at $N=2$}
Due to the existence of two conservation laws the system obtained from
(\ref{E:n3c1},\ref{E:n3c2}) at $N=1$ is of course trivial. However, at
$N=2$ we find that $d_1, d_2, r_1, r_2$ satisfy the 4 basic equations
\begin{eqnarray}
d_1{}'&=&d_1(d_1-d_2+r_2-r_1+2\beta_1),\label{E:r1}\\
d_2{}'&=&d_2(d_2-d_1+r_1-r_2+2\beta_2),\label{E:r2}\\
r_1{}'&=&d_2r_2-d_1r_1,\label{E:r3}\\
r_2{}'&=&d_1r_1-d_2r_2,\label{E:r4}
\end{eqnarray}
with $\beta_1 + \beta_2 = \eps$. We shall refer to this system as the
``symmetric form of $P_{III}$''. We now show how $P_{III}$ is obtained
from it.

Two first integrals were given in \eqref{E:P3conlawg}
\begin{equation}
d_1d_2=\kappa_1\kappa_2 e^{2 \eps x},\quad
r_1+r_2=\tfrac12(\rho_1+\rho_2)-\eps\label{E:P3ir}
\end{equation}
We resolve these by introducing new functions $g_i$ defined as follows:
\begin{equation}
  \label{eq:PIIIgsub}
  d_1 = \kappa_{1}\ g_1\ e^{\eps x} ,\quad 
  d_2 = \kappa_{2}\ g_1^{-1}\ e^{\eps x},\quad
  r_1= -g_1g_2+\tfrac12(\rho_1+\rho_2)-\eps,\quad
  r_2= g_1g_2.
\end{equation}
Then the equations
(\ref{E:r1}-\ref{E:r4}) take the form
\begin{eqnarray}
  g_1'&=&2g_1^2g_2+e^{\eps x}\kappa_{1}g_1^2-
  (\rho_2-\eps)g_1-e^{\eps x}\kappa_{2},\label{E:PIIg1}\\
  g_2'&=&-2g_1g_2^2-2e^{\eps x}\kappa_{1}g_1g_2+
  (\rho_2-\eps)g_2 + e^{\eps  x}\kappa_{1}(\tfrac12(\rho_1+\rho_2)-\eps)
,\label{E:PIIg2}
\end{eqnarray}
which is Hamiltonian with
\begin{equation}
H=g_1^2g_2^2+e^{\eps x}\kappa_{1}g_1^2g_2-
(\rho_2-\eps)g_1g_2-e^{\eps x}[\kappa_{1}(\tfrac12(\rho_1+\rho_2)-\eps)
g_1+\kappa_{2}g_2]
\end{equation}
(which is the same as the one given in \cite{O81,O87c,O99},
up to a simple transformation).

If we solve $g_2$ from \eqref{E:PIIg1}, substitute it into
\eqref{E:PIIg2}, and use the new variables
\begin{equation}
y(z)=\sqrt{\tfrac{\kappa_1}{\kappa_2}}\, g_1(x), \text{ where }z
=\tfrac{\sqrt{\kappa_1\kappa_2}}{\eps} e^{ \eps x},\label{E:P3ydef}
\end{equation}
 we get $P_{III}$
in the canonical form:
\begin{equation}
\ddy = \frac{1}{y}\left(\dy\right)^2 - \frac{1}{z}\dy  + y^3 + 
\frac1{z}\left({\tfrac{\rho_1}{\eps} y^2 - \tfrac{\rho_2}{\eps}}\right)
 - \frac1{y}.\label{E:P3}
\end{equation}

\subsection{Bilinear form of $P_{III}$}
The bilinear form of $P_{III}$ is basically nothing else but
\eqref{E:P3bilchaing} with periodic $\tau$'s. Since the $\tau$'s
appear in pairs ($\tau_1,\,\tau_2^+$ and $\tau_2,\,\tau_1^+$) and
since bilinear equations are gauge invariant, we propose the following
gauge transformation and scaling
\begin{equation}
\tau_1=\sqrt{\kappa_1}\,\bar\tau_1 e^{\theta x},\, \ 
\tau_2^+=\bar\tau_2^+e^{\theta x},\, \
\tau_1^+=\bar\tau_1^+,\, \
\tau_2=\sqrt{\kappa_2}\,\bar\tau_2,\ \ 
\theta=\tfrac14(\sigma_1-\sigma_2),\label{P3gauge}
\end{equation}
which yield, after a subsequent $x$-translation $x\to
x-\tfrac1{2\eps}\log(\kappa_1\kappa_2)$, the following novel bilinear
form of $P_{III}$:
\begin{eqnarray}
D_x \, \bar\tau_2^+ \cdot \bar\tau_1 &=& 
e^{\eps x}\,\bar\tau_1^+\bar\tau_2,\label{E:P3b1}\\
D_x \, \bar\tau_1^+ \cdot \bar\tau_2 &=& 
e^{\eps x}\,\bar\tau_2^+\bar\tau_1,\label{E:P3b2}\\
D_x^2\, \bar\tau_2^+ \cdot \bar\tau_1 &=& 
e^{\eps x}\,(D_x -\rho_1) \bar\tau_1^+ \cdot \bar\tau_2,\label{E:P3b3}\\
D_x^2\, \bar\tau_1^+ \cdot \bar\tau_2 &=& 
e^{\eps x}\,(D_x -\rho_2) \bar\tau_2^+ \cdot \bar\tau_1.\label{E:P3b4}
\end{eqnarray}
From this $P_{III}$ can also be derived directly: Comparing
(\ref{eq:PIIIgsub},\ref{E:dtprod},\ref{E:P3ydef}) (after the
aforementioned gauge transformation) one can solve for $\bar\tau_2$ in
terms of $y$ and the other $\bar\tau$'s. Then from \eqref{E:P3b1} one
solves for $\partial_x\bar\tau_1$ and from \eqref{E:P3b2} for
$\partial_x\bar\tau_2^+$. Then \eqref{E:P3b3}$\times y-$\eqref{E:P3b4}
is nothing but the $P_{III}$ equation \eqref{E:P3}.  In these new
variables $y$ and $z$ we also have (cf. (\ref{ytau4},\ref{ytau5}))
\begin{equation}
y(z)=\partial_z\log\frac{\bar\tau_2^+}{\bar\tau_1}.\label{ytau3}
\end{equation}

It should be noted that bilinear forms are sensitive to changes in the
independent variable. For example, if we express the
\hyphenation{bi-linear}bilinear system obtained after the
transformation \eqref{P3gauge} (but without the $x$-translation), in
terms of the $z$-coordinate (now regarding the tau-functions
$\bar\tau_2^+, \tau_1, \bar\tau_1^+, \tau_2$ as functions of $z$) we
get a system of equations which is no longer expressible in Hirota
$D$-operators only as it also involves ordinary $z$-derivatives.
However, the system so obtained can be shown to be a reduction of an
integrable system contained in the so called ``modified 2-component KP
hierarchy'' (at least in the restricted case where $\rho_2 = \pm
\rho_1$) and hence the functions $\bar\tau_2^+, \tau_1, \bar\tau_1^+,
\tau_2$ introduced here can be thought of as genuine tau-functions
in the sense of Sato theory.

\subsection{Lax pair for $P_{III}$}
Just as for $P_{IV}$, a Lax pair for the
Painlev\'e III equation can be derived from the linear formulation
(\ref{E:Mlin},\ref{E:Ldlin}) introduced earlier for the general chain
equations \eqref{E:chaing}.

As before, the standard periodic closing \eqref{E:PIVfieper} involves
a shift in the eigenvalues $\nu_j$ associated with the eigenfunctions
$\fie_j$ (for the $P_{III}$ case one has $\nu_3 = \nu_1 - \eps$).
Hence, from the condition $\fie_3(\nu_3) = \fie_1(\nu_1)$ and
expression \eqref{E:PIIIf} it follows that $f_3 = f_1 + \eps$.
Furthermore, due to the relation $u_3(\lambda) = u_1(\lambda + \eps)$
there will exist eigenfunctions of the Schr\"odinger equation
\eqref{E:eig1} for which $\psi_3(\lambda,x) = \psi_1(\lambda+\eps,x)$
holds. This then, bearing in mind the relation \eqref{E:PIIIF} and the
fact that $A_3 = A_1$ (actually $A_3^2 = A_1^2$ ; we choose the sign
$A_3 = A_1$), yields the following periodic closing of the
linear equations \eqref{E:Mlin}
\begin{eqnarray}
\partial_x \psi_1  &=&  A_1^{-1} \psi_2 + (\lambda + f_1) 
\psi_1\label{Lax3.1.1}\\
\partial_x \psi_2  &=&  A_2^{-1} \psi_1^s + (\lambda + f_2) 
\psi_2,\label{Lax3.1.2}
\end{eqnarray}
 and \eqref{E:Ldlin}
\begin{eqnarray}
(A_1 A_2)^{-1} \psi_1^s + A_1^{-1} (2\lambda + f_1 + f_2 - 
(\log A_1)^\prime) \psi_2 + 2 (\lambda - \nu_1) A_1^{-2} \psi_1 &=& 
0,\label{Lax3.2.1}\\
(A_1 A_2)^{-1} \psi_2^s + A_2^{-1} (2\lambda + \eps + f_1 + f_2 - 
(\log A_2)^\prime) \psi_1^s + 2 (\lambda - \nu_2) A_2^{-2} \psi_2 &=& 
0,\label{Lax3.2.2}
\end{eqnarray}
where $\psi_i^s$ stands for $\psi_i(\lambda+\eps,x)$.

It is now advantageous to change to scaled eigenfunctions
\[
\wpsi_i:=A_i^{-1/2}e^{(-\varepsilon -2(\nu_1+\nu_2)+r_1+r_2)x/4}\,\psi_i
\]
(recall that $r_1+r_2$ is a constant). After also changing to the new
independent variable $\specz=\varepsilon
z=\sqrt{\kappa_1\kappa_2}e^{\varepsilon x}$ and using the previously
obtained formulae we can write (\ref{Lax3.1.1}-\ref{Lax3.2.2}) as
\begin{eqnarray}
\varepsilon\specz\partial_\specz \wpsi_1&=&
\tfrac14(4\lambda+d_1+d_2+\varepsilon-4\nu_1+4\nu_2+2r_1-2r_2)\wpsi_1
+\sqrt{\specz}\wpsi_2,\\
\varepsilon\specz\partial_\specz \wpsi_2&=&
\tfrac14(4\lambda+d_1+d_2-\varepsilon+4\nu_1-4\nu_2-2r_1+2r_2)\wpsi_2
+\sqrt{\specz}\wpsi_1,
\end{eqnarray}
\begin{eqnarray}
\specz \wpsi_1^s+\sqrt{\specz}(2\lambda+d_1+r_2-2\nu_2)\wpsi_2+
2d_1(\lambda-\nu_1)\wpsi_1&=&0,\\
\specz \wpsi_2^s+\sqrt{\specz}(2\lambda+d_2+r_1-2\nu_1+2\varepsilon)\wpsi_1^s+
2d_2(\lambda-\nu_2)\wpsi_2&=&0.
\end{eqnarray}

In exactly the same way as for the $P_{IV}$ equation, this system of
difference equations (in the spectral parameter) can be cast into a
more standard form by introducing the formal Fourier transform of the
eigenfunctions $\psi_j(\lambda,x)$:
\begin{equation}
\tilde\Phi_j(k,x) := \int\ d\lambda\ e^{i k \lambda}\ 
\wpsi_j(\lambda,x).
\end{equation}
In terms of the new variable $\xi = \exp( -i \eps k)$ and the new
dependent variables $\Phi_j(\xi,\specz) := \tilde\Phi_j(k,x)$, we then
obtain the linear systems
\begin{eqnarray}
\varepsilon\xi\partial_\xi
\begin{pmatrix}\Phi_1\\\Phi_2\end{pmatrix}
&=&\frac12\left[D_1+ A + \frac1{\xi-\specz}B
\right]\begin{pmatrix}\Phi_1\\\Phi_2\end{pmatrix},\label{P3lax1}\\
\varepsilon\specz\partial_\specz
\begin{pmatrix}\Phi_1\\\Phi_2\end{pmatrix}
&=&\frac12\left[D_2+A-\frac1{\xi-\specz}B
\right]\begin{pmatrix}\Phi_1\\\Phi_2\end{pmatrix},
\end{eqnarray}
where
\begin{equation}
A=\begin{pmatrix}\nu_2-\nu_1 & \sqrt\specz\\
\xi\sqrt\specz & \nu_1-\nu_2 \end{pmatrix},\quad
B=\begin{pmatrix}\xi r_1 & -d_2r_2/\sqrt{\specz}\\
-\xi d_1r_1/\sqrt{\specz} &\xi r_2  \end{pmatrix},
\end{equation}
\begin{equation}
D_1=-(\nu_1+\nu_2)
(\begin{smallmatrix}1 & 0\\0 & 1\end{smallmatrix}),\quad
D_2=[\tfrac12(d_1+d_2)+\nu_1+\nu_2]
(\begin{smallmatrix}1 & 0\\0 & 1\end{smallmatrix})
+[\tfrac12\varepsilon+r_1-r_2]
(\begin{smallmatrix}1 & 0\\0 & -1\end{smallmatrix}).\label{P3lax4}
\end{equation}
The compatibility of these two matrix equations yields the equations
(\ref{E:r1}-\ref{E:r4}).

\section{Conclusions}
In this paper we have constructed Darboux chains from (scalar)
Schr\"odinger equations for generic second order energy-dependent
potentials and then proceeded to discuss periodic reductions of such
chains. The Darboux chains were classified into three types, the so
called $P_{IV-V}$, $P_{III}$ and $P_{VI}$ branches, which derive their
names from the Painlev\'e equations that arise as the lowest period
(non-trivial) reductions contained in each of them. (A detailed
discussion of the $P_{VI}$ branch will be given in a sequel to the
present paper.)

We described in detail the construction of the generic chain equations
(\ref{E:PIVchain}) and (\ref{E:n3c1},\ref{E:n3c2}) for the $P_{IV-V}$
and $P_{III}$ branches, and their bilinearization
(\ref{E:bilchaing},\ref{E:P3bilchaing}). For the periodically closed
chains we described the reduction to the corresponding Painlev\'e
equations, with an intermediate equation in Hamiltonian form (see also
\cite{S99}). 

The bilinearization of the generic Darboux chains in the $P_{IV-V}$
and $P_{III}$ branches led to a representation of the eigenfunctions
in terms of tau-functions (\ref{E:PIVbiltra},\ref{E:fietilfietra}),
which in reduction gave rise to bilinear representations and
tau-function formulae for the $P_{III-V}$ equations
(\ref{E:PIVbilfo},\ref{E:PVbilfo},\ref{E:P3b1}-\ref{E:P3b4}) and their
solutions (\ref{ytau4},\ref{ytau5},\ref{ytau3}). From the bilinear
form of the $P_{IV-V}$ branch it is clear that the equations described
by it are identical to the so called $A_n^{(1)}$-type dynamical
systems introduced by Noumi and Yamada \cite{NY98a}. The exact nature
of the tau-functions associated to the periodic reductions of this
branch in the context of Sato theory will be discussed in a
forthcoming paper (see \cite{W03} for a discussion of the $P_{IV}$
case). The precise link between the tau-functions that appear in the
periodic reductions of the $P_{III}$ branch and Sato theory is
currently being investigated.

The construction of the Darboux chains presented in this paper also
allowed us to systematically derive Lax representations for the chain
equations and their reductions, resulting in a novel Lax description
of the $P_{III}$ equation (\ref{P3lax1}-\ref{P3lax4}). The Lax pairs
obtained for the $P_{IV-V}$ branch were again related to those
obtained by Noumi and Yamada for the $A_n^{(1)}$-type
dynamical systems mentioned before.

The interpretation of the Hamiltonians associated with the periodic
reductions of the Darboux chains, in terms of the tau-functions that
describe their solutions, as well as a detailed investigation of the
higher order members -- i.e., those corresponding to chains with higher
periods -- in the different branches of chain equations, are topics
that will be addressed in subsequent papers.

\section*{Acknowledgments}
The authors would like to thank Prof. A. Veselov for discussions.  J.H.
would like to thank J. Satsuma for warm hospitality during a visit to
the Tokyo University Graduate School of Mathematical Sciences. The
authors have also benefited from discussions and correspondence with R.
Conte, M. Noumi and K. Takasaki. This work was partially supported by a
grant-in-aid from the Japan Society for the Promotion of Science
(JSPS).

\end{document}